\documentclass[preprint,superscriptaddress,nofootinbib]{revtex4-1}
\usepackage{graphicx}
\usepackage{dcolumn}
\usepackage{bm}
\usepackage{amsmath}
\usepackage{amsfonts} 
\usepackage{amssymb}
\usepackage{latexsym}
\usepackage{bbm}
\usepackage{color}
\usepackage{amssymb}
\usepackage{comment}
\usepackage{amsthm}
\usepackage{epsf}
\usepackage{epsfig}
\usepackage{caption3}
\usepackage{appendix}
\usepackage{cases}
\usepackage{subfigure}
\usepackage{multirow}
\usepackage{float}
\usepackage{makecell}

\usepackage{booktabs}
\usepackage{threeparttable}
\usepackage{mathtools}
\usepackage{diagbox}

\newcommand{\beq}{\begin {equation}}
\newcommand{\eeq}{\end   {equation}}
\newcommand{\bea}{\begin {eqnarray}}
\newcommand{\eea}{\end   {eqnarray}}
\newcommand{\baa}{\begin {array}   }
\newcommand{\eaa}{\end   {array}   }
\newcommand{\bit}{\begin {itemize} }
\newcommand{\eit}{\end   {itemize} }
\newcommand{\be }{\begin {equation}}
\newcommand{\ee }{\end   {equation}}

\newcommand{\zboson}{\text{Z}}
\newcommand{\wpm}{W^{\pm}}

\newcommand{\lh}{h}

\newcommand{\sinbma}{\text{sin}(\beta-\alpha)}
\newcommand{\tanb}{\text{tan}(\beta)}

\newcommand{\hboson}{{h}}
\newcommand{\Hboson}{{H}}
\newcommand{\Aboson}{{A}}
\newcommand{\Hpm}{{H}^{\pm}}
\newcommand{\mh}{M_{{h}}}
\newcommand{\mH}{M_{{H}}}
\newcommand{\mA}{M_{{A}}}
\newcommand{\mcH}{M_{{H}^{\pm}}} 

\newcommand{\gev}{\text{GeV}}

\newcommand{\threefbm}{\text{300}~\text{fb}^{-1}}

\newcommand{\cmsfourteen}{\sqrt{s}=~\text{14 TeV}}

\begin{document}

\title{Analysis of the $gg\to H\to hh\to4\tau$ process \\ in the 2HDM lepton specific model at the LHC}
\author{Yan Ma}
\email[]{20214016005@mails.imnu.edu.cn}
\affiliation{\scriptsize College of Physics and Electronic Information,
Inner Mongolia Normal University, Hohhot 010022, PR China}

\author{A. Arhrib }
\email[]{\small a.arhrib@gmail.com}
\affiliation{\scriptsize Abdelmalek Essaadi University, Faculty of Sciences and Techniques,
B.P. 2117 Tétouan, Tanger, Morocco \& \\
 Department of Physics and Center for Theory and Computation, National Tsing Hua
University, Hsinchu, Taiwan 300}

\author{S. Moretti}
\email[]{\small s.moretti@soton.ac.uk; stefano.moretti@physics.uu.se}
\affiliation{\scriptsize School of Physics \& Astronomy, University of Southampton,
Southampton SO17 1BJ, UK \& \\
Department of Physics \& Astronomy, Uppsala University, Box 516, 75120 Uppsala, Sweden}

\author{S. Semlali}
\email[]{\small s.semlali@soton.ac.uk}
\affiliation{\scriptsize School of Physics and Astronomy, University of Southampton,
Southampton SO17 1BJ, UK \& \\
Particle Physics Department, Rutherford Appleton Laboratory, Chilton, Didcot, Oxon OX11 0QX,
United Kingdom}

\author{Y. Wang}
\email[]{\small wangyan@imnu.edu.cn}
\affiliation{\scriptsize College of Physics and Electronic Information,
Inner Mongolia Normal University, Hohhot, 010022, China
\& \\
Key Laboratory for Physics and Chemistry of Functional Materials,
Inner Mongolia Normal University, Hohhot, 010022, China}

\author{Q.S. Yan}
\email[]{\small yanqishu@ucas.ac.cn}
\affiliation{\scriptsize Center for Future High Energy Physics, Chinese Academy of Sciences,
Beijing 100049, PR China \& \\ School of Physics Sciences, University of Chinese Academy of Sciences,
Beijing 100039, PR China}

\vspace*{-1cm}
\begin{abstract}
{\footnotesize\baselineskip=15pt
We analyse the signature of a light Higgs boson pair in the 2-Higgs Doublet Model(2HDM) Type-X (or lepton specific) over the  parameter spaces allowed by theoretical self-consistency requirements as well as the latest experimental constraints from the Large Hadron Collider (LHC),  precision test data and $B$ physics. Over the viable regions of the latter, wherein the Standard Model (SM)-like Higgs boson discovered at the LHC in 2012 is the heavier CP-even state of the 2HDM, $H$, it is found that the SM-like Higgs boson can decay into a pair of the lighter CP-even Higgs boson, $h$, via the process $H\to hh$ with a Branching Ratio (BR) of $5\%-10\%$ or so, (with $2 m_h < m_H =125$ GeV). Furthermore, in the Type-X scenario, the lighter Higgs bosons $h$ can dominantly decay into two $\tau$'s due to a large $\tan\beta$. Therefore, the pair of lighter Higgs bosons can altogether decay into a 4 $\tau$ final state. In order to suppress the huge SM background events, we confine ourself to consider the fraction of signal events with two Same-Sign (SS) $\tau$'s further decaying into same sign leptons while the other two $\tau$'s decay hadronically. By using Monte Carlo (MC) and Machine Learning (ML) tools, we thus focus on the analysis of the signal process $pp\to H\to hh\to \tau^{+}\tau^{-}\tau^{+}\tau^{-}\to \ell v_\ell \ell v_\ell \tau_h \tau_h$ (where $\ell= e, \mu$ and $\tau_h$ means a hadronic decay of the $\tau$) and explore the feasibility of such a search at the LHC for a collision energy $\cmsfourteen$ and a  luminosity $\threefbm$.

}
\end{abstract}

\maketitle
\tableofcontents

\section{Introduction}
The LHC has played an irreplaceable role in our understanding of the microcosmic subatomic structure, particularly with the discovery of a 125 GeV scalar particle by the ATLAS and CMS experimental teams in 2012~\cite{Aad:2012tfa, Chatrchyan:2012ufa}. This was the last particle predicted by the Standard Model (SM) and its properties measured in the above experiments are currently consistent with the particle predicted by the SM {at the $5\sigma$ level}~\cite{ATLAS:2022vkf, CMS:2022dwd}. To investigate the Higgs sector at the LHC, wherein the presence of additional (pseudo)scalar particles might exist, remains an intriguing task, particularly in new physics models with additional doublet  fields. Therefore, to discover extra Higgs bosons at the LHC is one of the prime targets of Beyond the SM (BSM) signal searches.

One of the simplest extensions of the SM is the 2HDM, which contains two complex Higgs doublets~\cite{Lee:1973iz,Deshpande:1977rw,Branco:2011iw}. After Electro-Weak Symmetry Breaking (EWSB), there are three Goldstone bosons, which are ‘eaten’ by the $\wpm$ and $\zboson$ bosons while the remaining five degrees of freedom incarnate into physical Higgs bosons: two neutral CP-even scalars ($\hboson$ and $\Hboson$, with $\mh \textless \mH$), one CP-odd pseudoscalar ($\Aboson$) and two charged Higgs states $\Hpm$ (with mixed CP status).

In order to be consistent with the stringent experimental constraints from Flavour Changing Neutral Currents (FCNCs) at the tree level, typically, a $Z_2$ symmetry is introduced into the Yukawa sector such that each type of fermion only couples to one of the doublets of the 2HDM~\cite{Glashow:1976nt}. Depending on the $Z_2$ charge assignments of the Higgs doublets, we can define four basic scenarios, known as (Yukawa) Types. In the lepton specific scenario (or Type-X), the extra Higgs doublet $\phi_2$ only couples to the lepton sector and not to the quark sector, where the light Higgs boson mainly decays to $\tau\tau$ for the wide range of parameter spaces~\cite{Aoki:2009ha}.


There have been several experiments for searching the SM Higgs exotic decays through four fermion final states $pp \to H \to XX \to f_1 f_2 f_3 f_4$. Such process benefits from the large cross-section of $\sigma(pp\to H)$, and the branching ratio of the SM-like Higgs exotic decay $Br(H\to hh)$ could reach up to 10\% in some BSM models.  Motivated by the recent searches for $\tau\tau\tau\tau$ by CMS experiments~\cite{CMS:2015twz,CMS:2017dmg,CMS:2019spf}, we will study the process $pp\to H\to hh \tau\tau\tau\tau$ in the 2HDM type-X model. In such a scenario, $h \to \tau^+\tau^-$ decay could be enhanced, which can lead  to a sizable $4\tau$ final state.  Therefore, it is worth to examine whether this statement is robust enough after taking into account the effects of parton shower, hadronization, heavy flavor decays and detector resolution.

In this paper,  we assume that the heavier CP-even Higgs boson $H$ is the observed SM-like Higgs boson whose properties agree with the LHC measurements.  In such a parameter space, a heavier Higgs boson can decay into two light Higgs bosons $h$ when kinematically allowed, i.e., $2 \mh < 125~\gev$. We perform a detailed MC study on the signal process $pp\to H\to hh\to 4\tau \to lv_llv_l\tau_h\tau_h $~(with $l = e, \mu$ and $\tau_h$ ) and relevant background processes, in order to examine its feasibility at the LHC. 
We do so after taking into account relevant theoretical and experimental conditions. In order to improve the experimental significance, we will require two leptons to be the same sign (SS) leptons. Our analysis demonstrates that, by using this signature, the relevant parameter space can be either discovered or ruled out by using the  current integrated luminosity at the LHC. 


The paper is organised as follows. In the next section, we briefly describe the 2HDM and its Yukawa scenarios, then introduce a few Benchmark Points (BPs) for our MC analysis which pass all present theoretical and experimental constraints. In the following section, we perform a detailed collider analysis of these BPs and examine the potential to discover the aforementioned signature within the 2HDM Type-X scenario. Finally, we present some conclusions.

\section{The 2HDM}
The scalar sector of the 2HDM consists of two weak isospin doublets with hyper-charge $Y = 1$. The most general  Higgs  potential for the 2HDM that complies with the $\rm{SU(2)_L \times U(1)_Y}$ gauge structure of the EW sector of the SM has the following form  \cite{Branco:2011iw}:
\begin{eqnarray}
V(\phi_1,\phi_2) &=& m_{11}^2(\phi_1^\dagger\phi_1) +
m_{22}^2(\phi_2^\dagger\phi_2) -
[ m_{12}^2(\phi_1^\dagger\phi_2)+\text{h.c.}] ~\nonumber\\&& 
+ \frac12\lambda_1(\phi_1^\dagger\phi_1)^2 +
\frac12\lambda_2(\phi_2^\dagger\phi_2)^2 +
\lambda_3(\phi_1^\dagger\phi_1)(\phi_2^\dagger\phi_2)+ 
\lambda_4(\phi_1^\dagger\phi_2)(\phi_2^\dagger\phi_1) ~\nonumber\\ && +
\frac12\left[\lambda_5(\phi_1^\dagger\phi_2)^2 +\rm{h.c.}\right]
+~\left\{\left[\lambda_6(\phi_1^\dagger\phi_1)+\lambda_7(\phi_2^\dagger\phi_2)\right]
(\phi_1^\dagger\phi_2)+\rm{h.c.}\right\},
\label{CTHDMpot}
\end{eqnarray}
where $\phi_{1}$ and $\phi_2$ are the two Higgs doublet fields. By hermiticity of such a  potential,  $\lambda_{1,2,3,4}$ as well as $m_{11,22}^2$ are   real parameters while $\lambda_{5,6,7}$ and $m_{12}^2$  can be complex, in turn enabling possible Charge and Parity (CP) violation effects in the Higgs  sector.
Upon enforcing two minimization conditions of the potential, $m^2_{11}$ and $m^2_{22}$ can be replaced by $v_{1,2}$, which are the Vacuum Expectation Values (VEVs) of the Higgs doublets $\phi_{1,2}$, respectively. 
Moreover, the coupling $\lambda_{1,2,3,4,5}$ can be substituted by the four physical Higgs masses (i.e., $\mh, \mH, \mA$ and $\mcH$) and the parameter $\sinbma$, where $\alpha$ and $\beta$ are, respectively, the mixing angles between CP-even  and the angle related to the VEV values, i.e., $\tan\beta=v_1/v_2$. Thus, the independent input  parameters can be taken as $\mh, \mH, \mA$ and $\mcH$, $\lambda_6$, $\lambda_7$, $\sinbma$, $\tanb$ and $m^2_{12}$.  

If both Higgs doublet fields of the general 2HDM couple to all fermions, the ensuing scenario can induce FCNCs in the Yukawa sector at tree level.  As intimated, to remedy this, a $Z_2$ symmetry is imposed  on the Lagrangian such that each fermion type interacts with only one of the Higgs doublets \cite{Glashow:1976nt}.  As a consequence, there are four possible types of 2HDM, namely Type-I, Type-II, Type-X (or lepton specific) and Type-Y (or flipped). However, such a symmetry is explicitly broken by the quartic couplings $\lambda_{6,7}$ and softly broken by the (squared) mass term $m^2_{12}$. 
 In what follows, we shall consider a {CP}-conserving (i.e., $m^2_{12}$ and $\lambda_5$ are real) 2HDM Type-X and assume that  $\lambda_{6} = \lambda_{7} = 0$ to forbid the explicit breaking of $Z_2$,
while also taking $m^2_{12}$ to be generally small, thereby preventing  large FCNCs at tree level, which are   incompatible with experiment. 

In general, the couplings of the neutral and charged Higgs bosons to fermions can be described by the Yukawa Lagrangian  given by \cite{Branco:2011iw} 
\begin{eqnarray}
- {\mathcal{L}}_{\rm Yukawa} = \sum_{f=u,d,l} \left(\frac{m_f}{v} \kappa_f^h \bar{f} f \lh + 
\frac{m_f}{v}\kappa_f^H \bar{f} f H 
- i \frac{m_f}{v} \kappa_f^A \bar{f} \gamma_5 f A \right) + \nonumber \\
\left(\frac{V_{ud}}{\sqrt{2} v} \bar{u} (m_u \kappa_u^A P_L +
m_d \kappa_d^A P_R) d H^+ + \frac{ m_l \kappa_l^A}{\sqrt{2} v} \bar{\nu}_L l_R H^+ + \text{h.c.} \right),
\label{Yukawa-1}
\end{eqnarray}
where $\kappa_f^S$ ($S=\lh,H$ and $A$) are the Yukawa couplings in the 2HDM, which are illustrated 
in Tab.~\ref{yuk_coupl} for the Type-X under consideration. Here, $V_{ud}$ refers to a Cabibbo-Kobayashi-Maskawa 
(CKM) matrix element and $P_{L,R}$ denote the left- and right-handed projection operators. The coupling of the two CP-even states $\lh$ and $H$ to gauge bosons $VV$ ($V = W^\pm, Z$) are proportional to $\sin(\beta-\alpha)$ and $\cos(\beta-\alpha)$, respectively. If we assume that either $\lh$ or $H$ can be the observed SM-like Higgs boson, the coupling  to gauge bosons is obtained for $\lh$ when $\cos(\beta-\alpha) \rightarrow 0$ and for $H$ when $\sin(\beta-\alpha) \rightarrow 0$. 
Therefore, each scenario can explain the 125 GeV Higgs signal at the LHC. 
However, following our works \cite{Arhrib:2021xmc,Wang:2021pxc,Arhrib:2021yqf,Li:2023btx}, though, we shall focus in the present paper on the scenario where $H$ mimics the observed signal with mass $\sim\,125$ GeV (as previously intimated).      

\begin{table}[H]
	\centering
	\renewcommand{\arraystretch}{1.2} %
	\setlength{\tabcolsep}{1.2pt}
	\begin{tabular}{c|c|c|c} 
		& $\kappa_u^{S}$ &  $\kappa_d^{S}$ &  $\kappa_\ell^{S}$  \\   \hline
		$\lh$~ 
		& ~ $  \cos\alpha/ \sin\beta$~
		& ~ $  \cos\alpha/ \sin\beta$~
		& ~ $  -\sin\alpha/ \cos\beta $~ \\
		$H$~
		& ~ $  \sin\alpha/ \sin\beta$~
		& ~ $  \sin\alpha/ \sin\beta$~
		& ~ $  \cos\alpha/ \cos\beta$~ \\
		$A$~  
		& ~ $  \cot \beta $~  
		& ~ $  -\cot \beta $~  
		& ~ $  \tan \beta $~  \\ 
	\end{tabular}
	\caption{Yukawa couplings of the fermions $f=u,d$ and $\ell$ to the neutral Higgs bosons $S=h,H$ and $A$ in the 2HDM Type-X.}
	\label{yuk_coupl}	
\end{table}	

\section{Bounds and constraints in the parameter space scans }\label{psc}
In order to identify regions that satisfy both theoretical requirements and experimental observations, the following numerical explorations of the possible parameter spaces have been studied: 
\begin{eqnarray}
&~& \mh \in [15, 60]~(\gev), \quad \mH = 125, \quad  \mA =~245(\gev), \quad  \mcH = 258~(\gev) \nonumber\\
&~& \sinbma \in [-0.25, -0.05], \quad \tanb \in [2, 13],\quad m_{12}^2\in [0,500] (\text{GeV})^2
\end{eqnarray}

The program \texttt{2HDMC-1.8.0}~\cite{Eriksson:2009ws} was used for calculating the cross sections ($\sigma$'s) and BRs as well as testing the following theoretical and experimental constraints.
\begin{itemize}
	\item Vacuum stability~\cite{Deshpande:1977rw}:  the scalar potential should be bounded by $\lambda_1 >0$, $\lambda_2>0$, $\lambda_3>-\sqrt{\lambda_1\lambda_2}$, $\lambda_3+\lambda_4-\lambda_5>-\sqrt{\lambda_1\lambda_2}$.
	\item Perturbativity constraints~\cite{Kanemura:1993hm, Akeroyd:2000wc}, which imply the condition $|\lambda_{i}|\textless 8\pi (i=,...5)$. 
 	\item Tree-level perturbative unitarity~\cite{Kanemura:1993hm, Arhrib:2000is, Akeroyd:2000wc}: requiring unitary when $2\to2$ scattering processes involving Higgs and gauge bosons at high energy.
    \item Oblique EW parameters ($S$, $T$ and $U$)~\cite{Peskin:1990zt, Peskin:1991sw}, which control the mass splitting between the Higgs states, with the following measured values~\cite{ParticleDataGroup:2022pth}:
\begin{align}
S= -0.02\pm 0.10,\quad T = 0.03\pm 0.12.
\end{align}
The correlation factor between $S$ and $T$ is set as $0.92$ for consistency at $95\%$ Confidence Level (CL). 
\end{itemize}

To account for potential additional Higgs bosons, exclusion bounds at $95\%$ CL  are enforced using the \texttt{HiggsBounds-5.9.0} program \cite{Bechtle:2020pkv}, which systematically checks each parameter point against the $95\%$ CL exclusion limits derived from Higgs boson searches conducted by LEP, Tevatron and LHC experiments. 

To ensure agreement with the measurements of the SM-like Higgs state, constraints are enforced using the \texttt{HiggsSignals-2.6.0} program \cite{Bechtle:2020uwn}, which incorporates the combined measurements of the SM-like Higgs boson from  LHC Run-1 and Run-2 data. 
	
Constraints from flavor physics are incorporated using  \texttt{SuperIso v4.1} \cite{Mahmoudi:2008tp} by utilizing the following observables: 
	\begin{itemize}
		\item BR$(B \to X_s \gamma) = (3.32 \pm 0.15) \times 10^{-4}$ \cite{HFLAV:2022esi},
  		\item	${\rm BR}(B^0\to \mu^+\mu^-)_{\text{(LHCb)}}$=$\left(1.2^{+0.8}_{-0.7}\right)\times 10^{-10}$~\cite{LHCb:2021awg,LHCb:2021vsc}, 
			\item ${\rm BR}(B^0\to \mu^+\mu^-)_{\text{(CMS)}}$=$\left(0.37^{+0.75}_{-0.67}\right)\times 10^{-10}$~\cite{CMS:2022mgd},
           \item	${\rm BR}(B_s\to \mu^+\mu^-)_{\text{(LHCb)}}$ = $\left(3.09^{+0.46}_{-0.43}\right)\times 10^{-9}$~\cite{LHCb:2021awg,LHCb:2021vsc},
			\item	${\rm BR}(B_s\to \mu^+\mu^-)_{\text{ (CMS)}}$=$\left(3.83^{+0.38}_{-0.36}\right)\times 10^{-9}$~\cite{CMS:2022mgd},
   \item BR$(B \to \tau \nu) = (1.09 \pm 0.20) \times 10^{-4}$ \cite{HFLAV:2022esi}. 
	\end{itemize}

\begin{figure}[h!]
	\hspace*{-0.3cm}
	\includegraphics[scale=0.35]{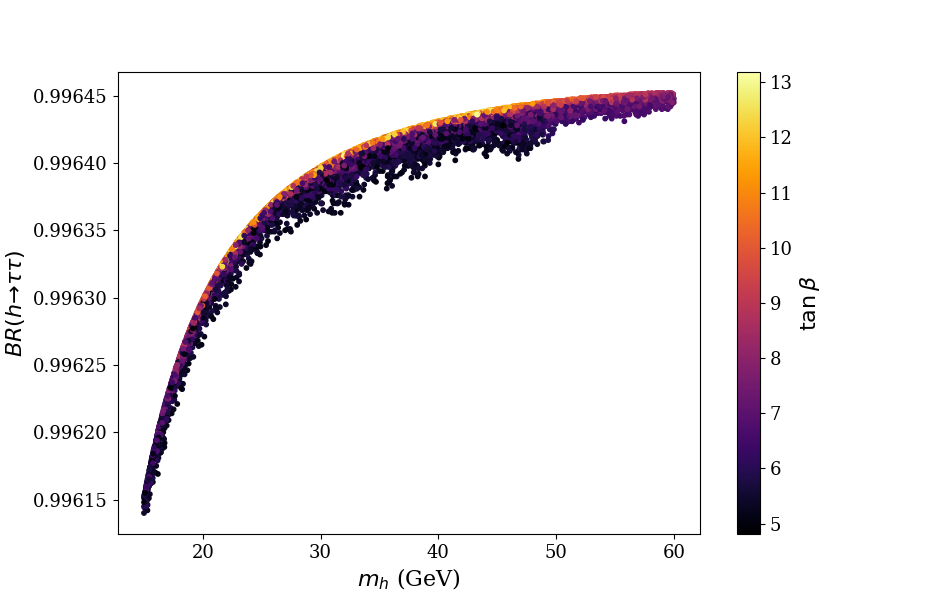}	
 \hspace*{-0.4cm}
	\includegraphics[scale=0.35]{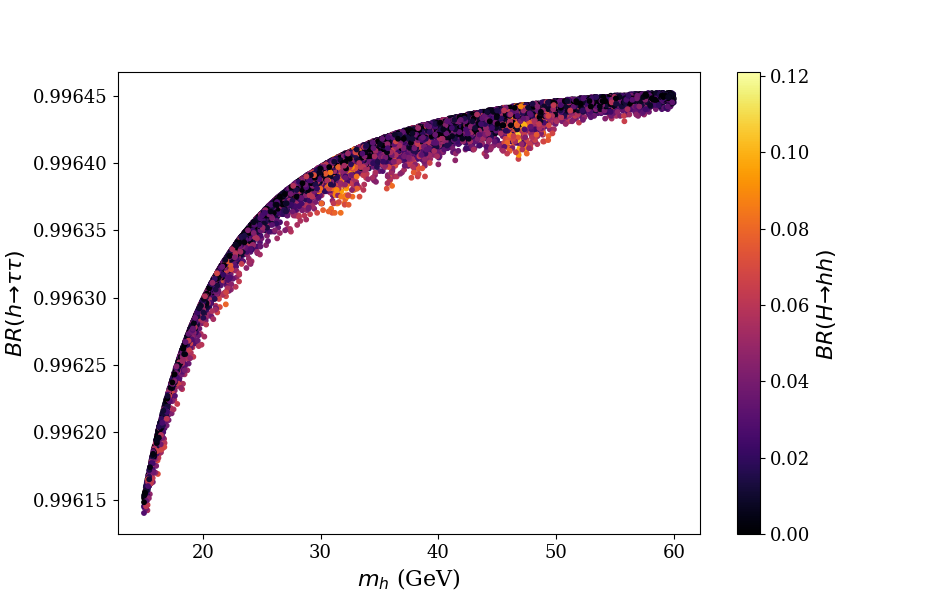}
	\caption{$\rm{BR}(h \to \tau\tau)$ as a function of $m_h$ vs. $\tan\beta$ (left panel) and $\rm{BR}(H\to hh)$ (right panel) in the 2HDM type-X.}
	\label{fig1}
\end{figure}

In the left panel of Fig.~\ref{fig1}, we show the branching ratio of the light Higgs decaying into pairs of $\tau$ leptons within the allowed parameter space of the 2HDM Type-X. The total width of the $h$ state ($\Gamma_h$) is obviously dominated by $h\to \tau \tau$. Such a significant branching ratio is attributed to the enhancement of the light Higgs couplings to leptons at large $\tan\beta$. The branching ratio of the SM-like Higgs boson decaying into a pair of light Higgs bosons is displayed on the right panel. The precise measurements of the Higgs boson couplings~\cite{ATLAS:2022vkf,CMS:2022dwd} tighten the constraints on Higgs decay to non-SM particles and force the $\rm{BR}(H\to hh)$ to be below 12\%. Both ATLAS and CMS collaborations have reported an upper limit of 12\%~\cite{ATLAS:2022vkf} and 16~\%\cite{CMS:2022dwd}, respectively,  on $\rm{BR}(H\to BSM)$ at 95\%CL.

To test the allowed parts of the parameter space, twelve BPs are given in Tab. \ref{t:BPs}, in terms of the 2HDM Type-X parameters entering our production and decay process, including the corresponding $pp \to H \to 2h \to 4\tau $ cross sections at
the LHC with $\sqrt{s} =14$ TeV. As one can see from such a table, the light Higgs boson mass varies from about $15~\gev$ to $60~\gev$, which can be paired produced from a $125~\gev$ Higgs boson. The cross sections of the BPs are around $2-5$ pb, which are large enough to be detected at the collider.

\begin{table}[h!]
	\begin{center}
			\begin{tabular}{|c| c| c| c| c|}
				\hline
				~ & $m_h$ & $\sin(\beta - \alpha)$ & $\tan\beta$ & $\sigma_{4\tau,~14~{\rm TeV}}$ (pb) \\
				\hline  
				BP1 &  19.04  & --0.12 & 8.06 & 2.53\\
                \hline 
				BP2 &  21.8 & --0.162 & 6.35 & 3.29\\
                \hline 
				BP3 &  25.11  & --0.16 & 5.82 & 2.18\\
                \hline 
				BP4 &  28.00  & --0.17 & 5.61 & 2.72\\
                \hline 
				BP5 &  32.14  & --0.15 & 6.36 & 4.01\\
                \hline 
				BP6 &  36.02  & --0.107 & 9.72 & 3.31\\
				\hline 
				BP7 &  44.00  & --0.188 & 5.10 & 2.43\\
                \hline 
				BP8 &  46.88  & --0.094 & 10.29 & 4.74\\
                \hline 
				BP9 &  47.30  & --0.10 & 9.92 & 3.88\\
                \hline  
				BP10 &  50.98 & --0.13 & 7.57 & 2.81\\
                \hline
				BP11 &  55.10 & --0.149 & 6.45 & 2.65\\
                \hline 
				BP12 &  58.16 & --0.11 & 8.91 & 1.80\\
				\hline  
			\end{tabular}
			\caption{2HDM Type-X input parameters and Leading Order (NLO) cross sections (at  parton level with $\sqrt{s} =14$ TeV) for each BP are presented. The unit of all masses is GeV and we fix the SM-like Higgs boson mass as $M_H$ = 125 GeV.
               }\label{t:BPs}
	\end{center}
\end{table}

In this paper, we mainly focus on two SS taus further leptonicly decay, while the other two tau hadronicly decay  (tau-jets). Thus, the signal will be SS leptons and two ($\tau$-)jets. The leptons are used for triggering purposes, so its kinematics is bound to comply with the trigger requirements.

\section{Collider phenomenology}
In this section, we present a detailed MC analysis for both signal and background events at the detector level. In the $pp\to H\to hh\to \tau^{+}\tau^{-}\tau^{+}\tau^{-}\to lv_l lv_l \tau_h \tau_h$ process, a light Higgs boson decays into two $\tau$ leptons and one of the $\tau$ decays into a charged lepton and neutrino, while the other $\tau$ decays hadronically 
(thus labelled as $\tau_h$). As a result, the final states consist of two SS leptons and two jets. The SM backgrounds are: $t \overline{t}\to l v_l l v_l b \overline{b}$, $W^\pm t b\to l v_l l v_l b \overline{b}$ (where the $W^\pm$ boson and $b$-jet are not from a top quark resonance), $W W j j\to l v_l l v_l j j)$, $Z j j\to l v_l l v_l j j$, $Z Z\to l v_l l v_l \tau_h \tau_h$, $t \overline{t} Z\to l v_l l v_l b \overline{b} \tau \tau$, $t \overline{t} Z Z\to l v_l l v_l l v_l b \overline{b} \tau \tau$ and $t \overline{t} W W\to l v_l l v_l l v_l j j $. 
In order to suppress the huge SM background process $Z j j$, we deliberately choose two SS leptons plus two hadronic $\tau$'s in the signal events, which is about $10\%$ of the total number of the $4\tau$ ones.


\subsection{Event generation and selection}
We use MadGraph5$_{-}$aMC@NLO v3.4.0~\cite{Alwall:2014hca} to compute the cross sections and generate both signal and background events at parton level.  The following kinematic cuts are adopted for the signal and background generations:
\begin{eqnarray}\label{cuts}
 |\eta (l,j)|< 2.5,~p_{\text{T}}(l,j)> 10~\gev,~E_T^{\text{miss}}>5~\text{GeV},~\Delta R(ll,lj,jj)>0.4.   
\end{eqnarray}

Then parton level events are then passed to Pythia~\cite{Sjostrand:2006za,Sjostrand:2014zea}  to simulate initial and final state radiation (i.e., the QED and QCD emissions), parton shower, hadronization and heavy flavor decays. We further use Delphes v3.4.2~\cite{deFavereau:2013fsa}   to simulate the detector effects. For each event, the anti-$k_t$ jet algorithm~\cite{Cacciari:2008gp}
is used to cluster jets with jet parameter $\Delta R = 0.4$ in the FastJet package~\cite{Cacciari:2011ma}.  The following kinematic cuts are adopted in order to emulate the detector acceptance:
\begin{eqnarray}\label{cuts}
 |\eta (l,j)|< 2.5,~p_T(l,j)> 10~\gev.   
\end{eqnarray}
The leptons are used for triggering purposes, so its kinematics is bound to comply with the
trigger requirements. \footnote{The CMS developed a double muon trigger to search for low pt muons from B meson decays that are close together. The same approach has been adopted recently to search for a pairs of electron from B meson decays, by tightening the topological selection on the two electrons L1 objects.  In our recent paper~\cite{Arhrib:2023apw}, we discussed the possibility of developing an electron-muon trigger to search for low pt leptons (pt~10 GeV) }

Then, we select two leptons and two jets final states from detector simulated events. In order to improve the signal event rate, the tau tagging is not applied. In order to further suppress the huge SM backgrounds, the SS leptons are selected. The cross sections for the signal and background processes with detector acceptance cuts, with $2l2j$ selection and with two SS lepton selection are shown in the  Tab.~\ref{t:CS_signal} and  in  Tab.~\ref{t:bg}, respectively.

\begin{table}[H]
	\begin{center}
			\begin{tabular}{|c| c| c| c| c| c| c| c| c| c| c| c| c|}
				\hline
			$\sigma$ (fb) & BP1  & BP2   & BP3   & BP4  & BP5   & BP6   & BP7   & BP8   & BP9   & BP10  & BP11   & BP12 \\
				\hline  
     parton level generation  & 8.95 & 11.37 & 7.38  & 9.05 & 13.32 & 11.33 & 10.43 & 
           23.16 & 19.26 & 15.81 & 17.00  & 12.75\\
                    \hline
           selection of 2l2j  & 0.67 &  0.88 & 0.59  & 0.76 &  1.10 &  0.96 &  0.88 &  1.97 &  1.62 &  1.46 &  1.70  &  1.38\\
                    \hline
           selection of SS 2l & 0.337 &  0.438 & 0.31 & 0.41 & 0.59 &  0.52 &  0.50 &  1.14 &  0.92 &  0.83 &  0.93  &  0.75\\
				\hline  
			\end{tabular}
			\caption{Cross sections for signals in the parton level, detector level, and after selection of SS leptons at $\cmsfourteen$ for $\threefbm$.
            }\label{t:CS_signal}
	\end{center}
\end{table}

\begin{table}[H]
	\begin{center}
			\begin{tabular}{|c| c| c| c| c| c| c| c| c|}
				\hline
				$\sigma$ (fb) & $t\bar{t}$ & $W^\pm tb$ & $W^+W^-jj$ & $Zjj$ & $ZZ$ & $t\bar{t}Z$ & $t\bar{t}ZZ$ & $t\bar{t}W^+W^-$ \\
				\hline
	    parton level generation&  16060 & 518.3 & 1053 & 317600 & 18.89 & 0.49 & $1.14\times10^{-4}$ & 0.02\\
				\hline
    	selection of 2l2j      &  8787.7 & 289.9 & 530.1 & 151086 & 10.0 & 0.33 & $1.1\times10^{-4}$ & 0.018\\
				\hline  
    	  selection of SS 2l     &  19.43 & 0.62 & 1.99 & 0 & 2.51 & 0.079 & $3.3\times10^{-5}$ & $7.6\times10^{-3}$\\
				\hline  
			\end{tabular}
			\caption{Background rates after the acceptance cuts in Eq.~(\ref{cuts})  at $\cmsfourteen$ for $\threefbm$.}\label{t:bg}
	\end{center}
\end{table}

\begin{figure}[ht]
\begin{minipage}{0.47\textwidth}
      \begin{center} 
                     (a)	\\
         \includegraphics[height=5.0cm]{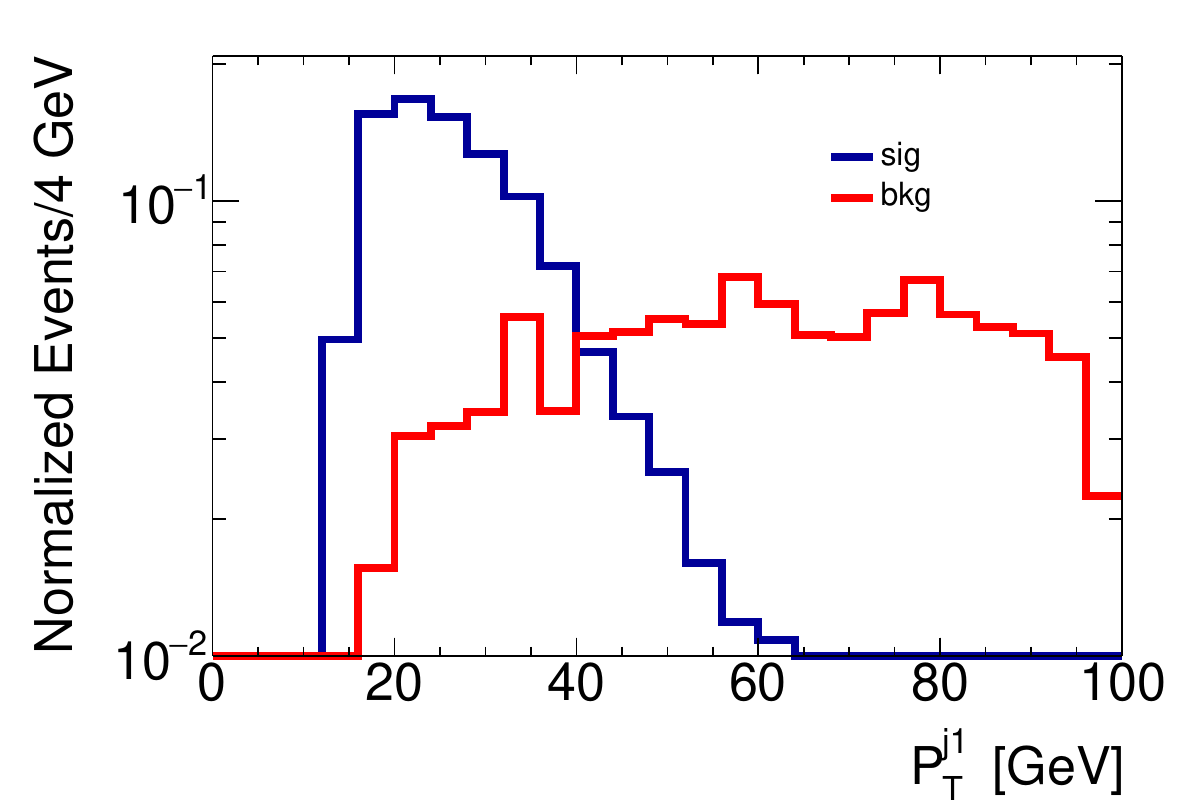}	
      \end{center}
\end{minipage}
\begin{minipage}{0.47\textwidth}
      \begin{center} 
                     (b)	\\
         \includegraphics[height=5.0cm]{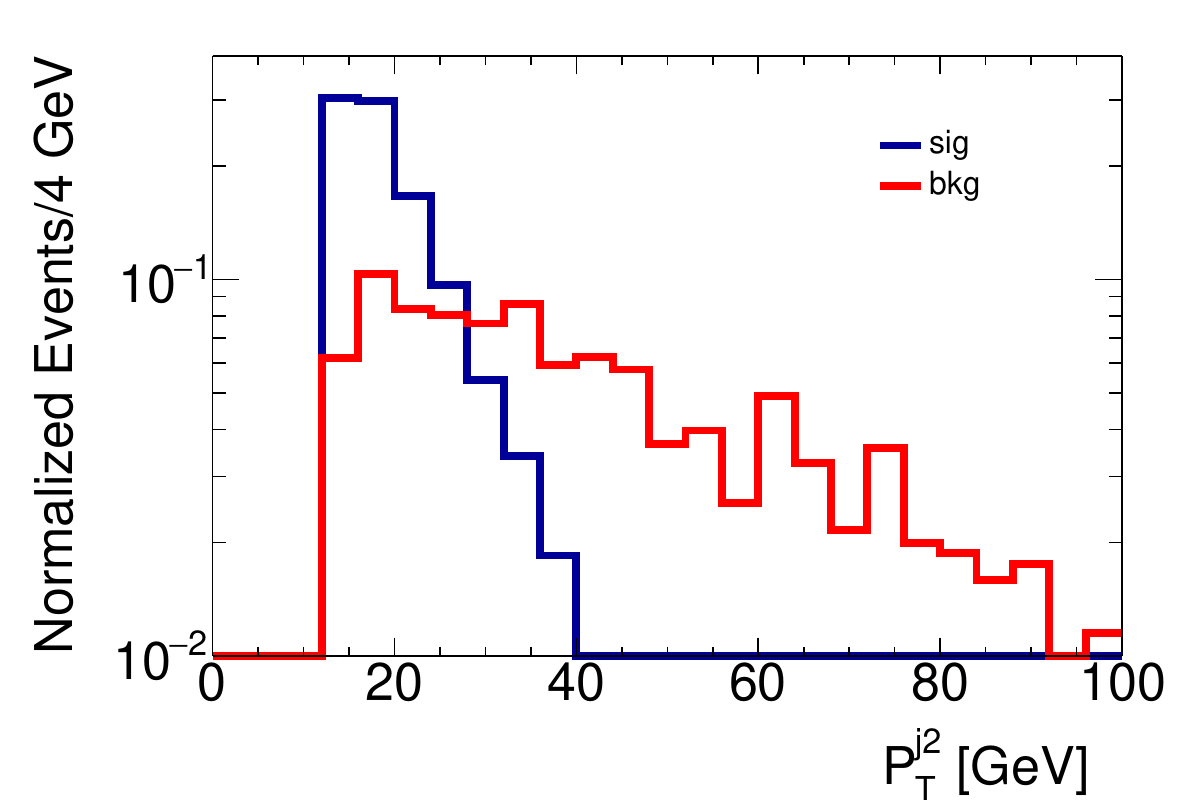}	
      \end{center}
\end{minipage}

\begin{minipage}{0.47\textwidth}
      \begin{center} 
                     (c)	\\
         \includegraphics[height=5.0cm]{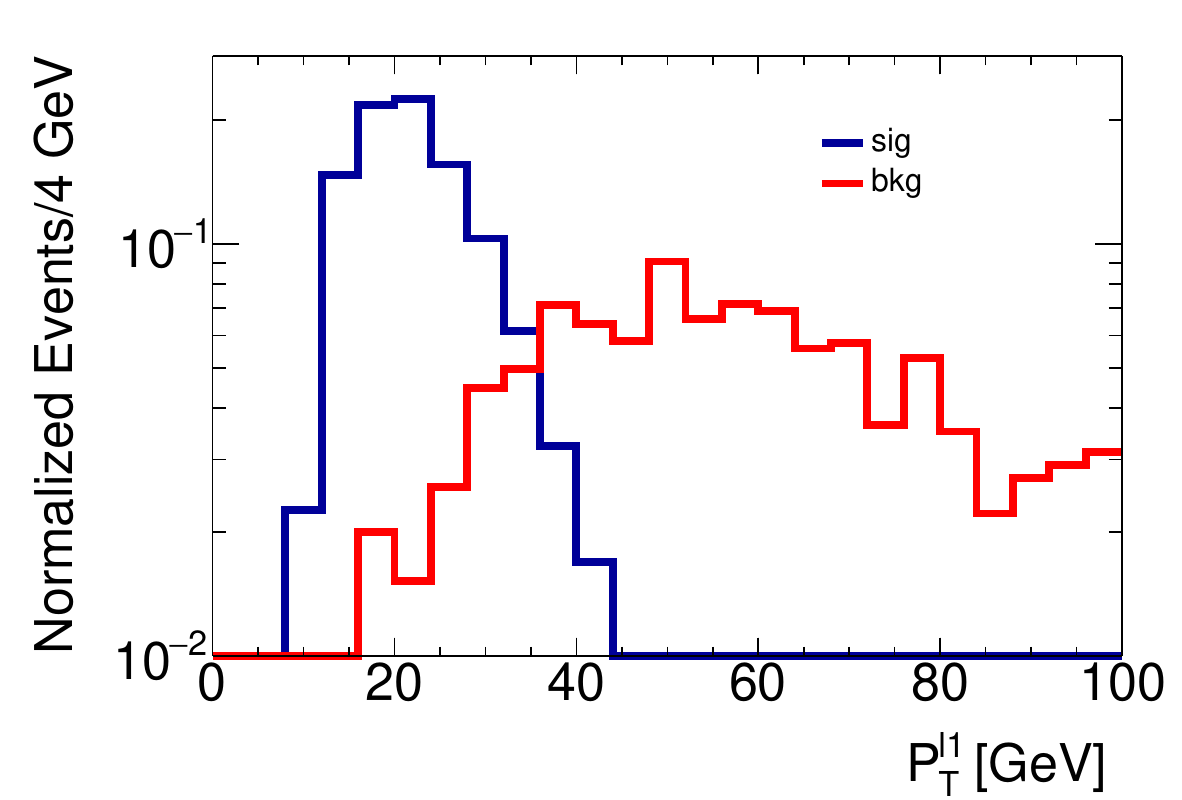}	
      \end{center}
\end{minipage}
\begin{minipage}{0.47\textwidth}
      \begin{center} 
                     (d)	\\
         \includegraphics[height=5.0cm]{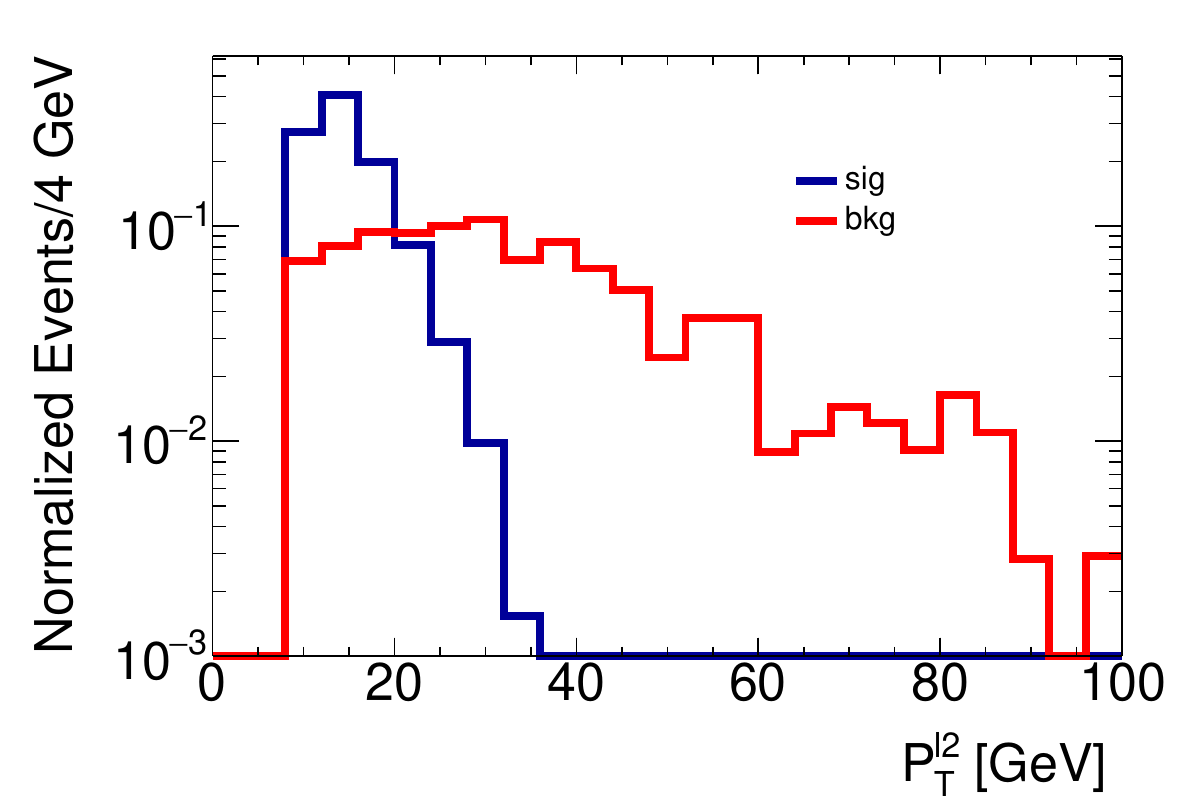}	
      \end{center}
\end{minipage}

\begin{minipage}{0.47\textwidth}
      \begin{center} 
                     (e)	\\
         \includegraphics[height=5.0cm]{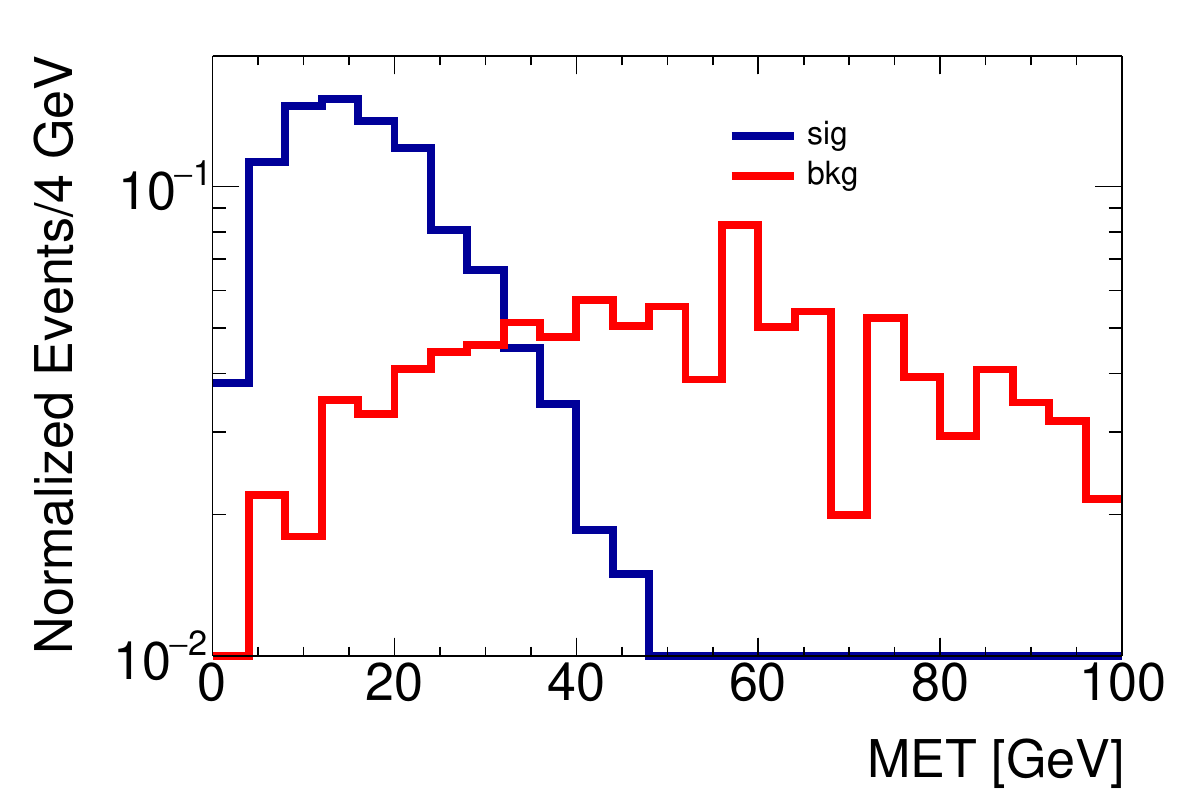}	
      \end{center}
\end{minipage}
 \caption{Distributions in (a)-(d) transverse momenta of the leading/second-leading jets/leptons and (e) MET for BP2 and backgrounds  when $\cmsfourteen$  and L = $\threefbm$.}\label{f:pt}
\end{figure}

In  Fig.~\ref{f:pt}, the transverse momentum distributions of the leading  and the second-leading jets and leptons as well as the transverse missing energy (MET) are presented for, e.g., BP2 and background events. It is clear that the transverse momenta of the signal final state are always much softer than for background events, because they emerge from a rather light Higgs bosons, which energies are capped at a mass of 125 GeV. 

\subsection{Event reconstruction}
To further improve the signal-to-background discrimination, we reconstruct some kinematic features of signal events. Specifically, 
because the light Higgs boson decays into two $\tau$ and one of the $\tau$'s  decays to a lepton and two neutrinos plus two $\tau$ neutrinos emerge from hadronic $\tau$ decays, there will be six neutrinos in a signal event. It is thus extremely hard to completely reconstruct the momentum of each light Higgs. 

\begin{figure}[ht!]
	\centering
	\begin{minipage}{8cm}
		\centering
		\includegraphics[width=8cm]{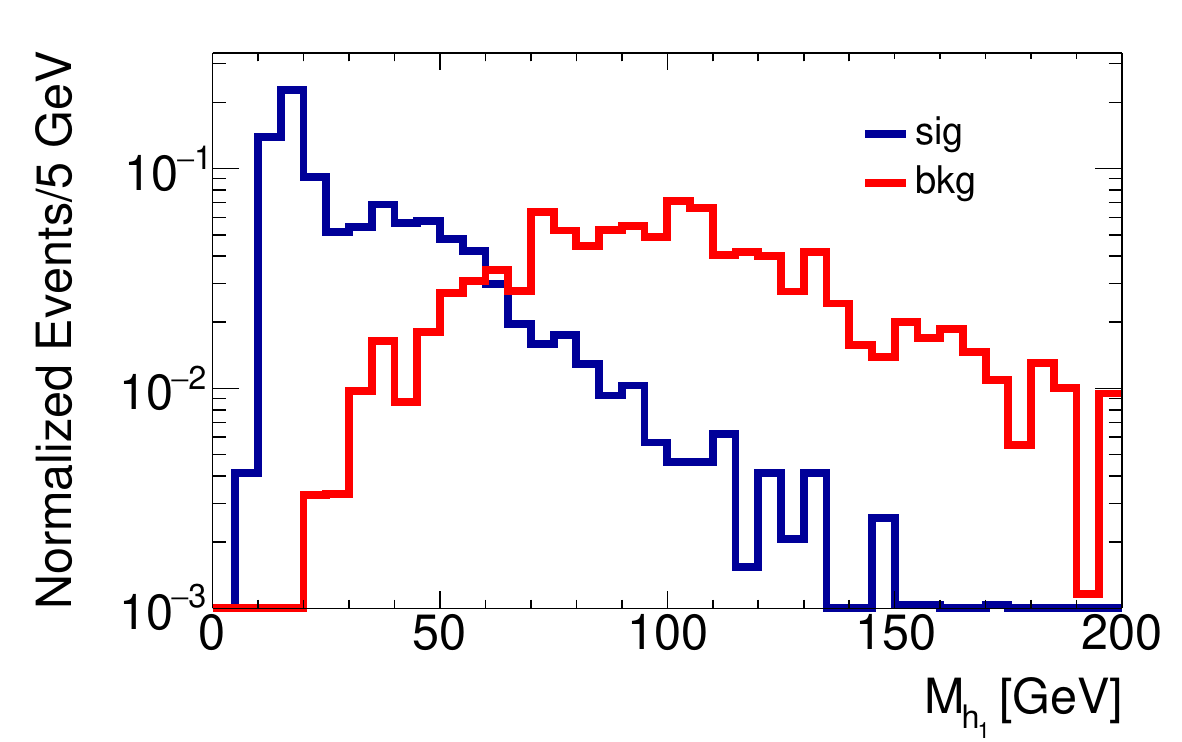}
	\end{minipage}
    \begin{minipage}{8cm}
		\centering
		\includegraphics[width=8cm]{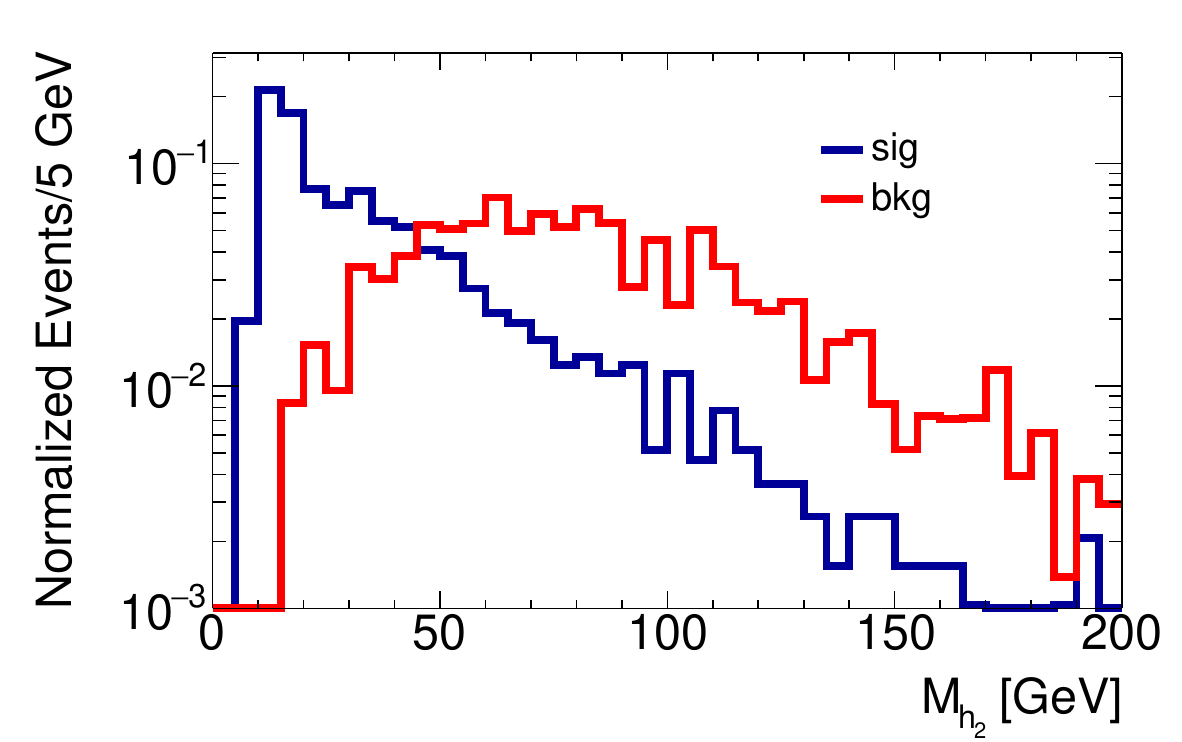}
	\end{minipage}
\caption{Distributions in invariant mass of a lepton and a jet for BP2 and backgrounds. The reconstructed $M_h$ values, corresponding to the best $\chi^2$ fit,  is labelled as $M_{lj}^1$ (left) and $M_{lj}^2$ (right), respectively,  when $\cmsfourteen$  and L = $\threefbm$.}\label{f:Mh1Mh2}
\end{figure}

However, since the lepton momenta and MET are all very small, we can approximately reconstruct the light Higgs boson with only one lepton and one jet. Thus, there are two possible combinations with two leptons and two $\tau$-jets in each event.  By defining a suitable $\chi^2$,
\begin{equation}\label{eq:h1h2}
\chi^2 = {(M_{{lj}}^1 - M_{{h}})}^2 + {(M_{lj}^2 - M_h)}^2,
\end{equation}
where $M_{h}$ is the light Higgs mass, we pair the two leptons and two jets to find a combination which minimizes it and then assign these to $M_{lj}^1$ and $M_{lj}^2$, respectively. Taking again BP2 as an example, the reconstructed light Higgs mass distributions are shown in Fig.~\ref{f:Mh1Mh2}.

\begin{figure}[ht!]
	\centering
	\begin{minipage}{8cm}
		\centering
		\includegraphics[width=8cm]{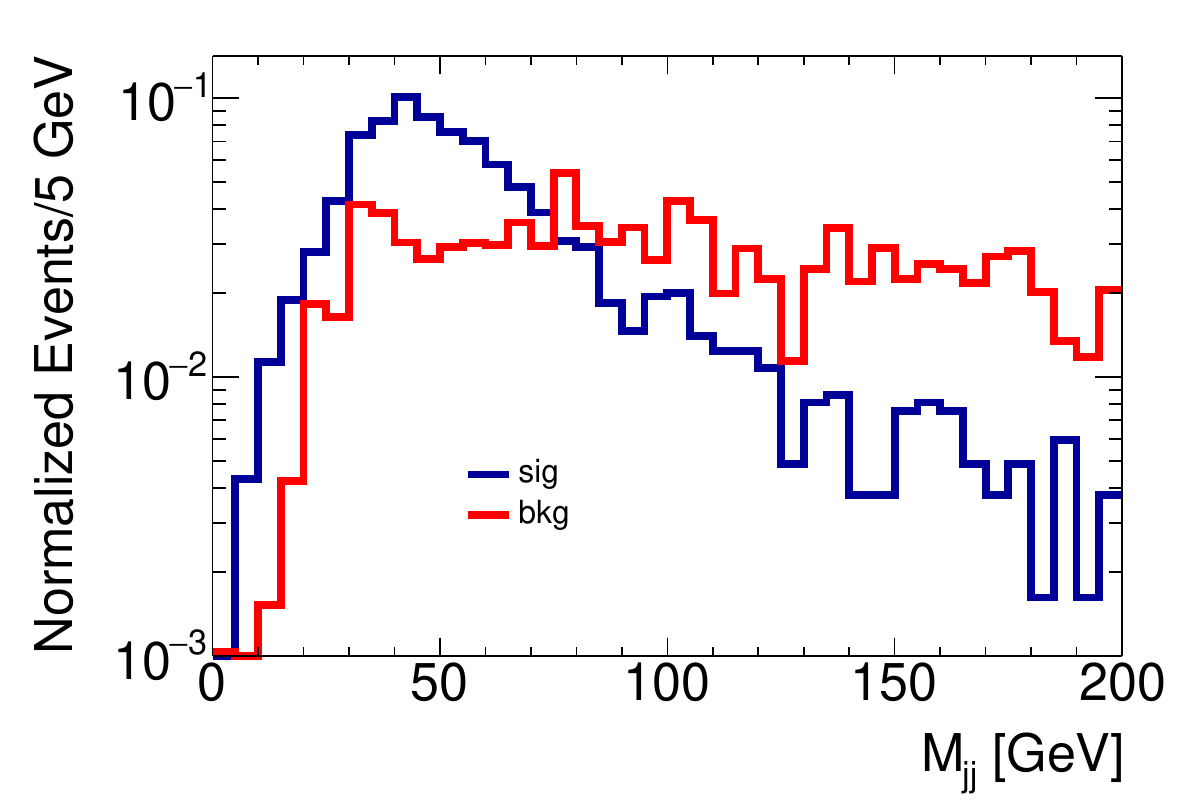}
	\end{minipage}
    \begin{minipage}{8cm}
		\centering
		\includegraphics[width=8cm]{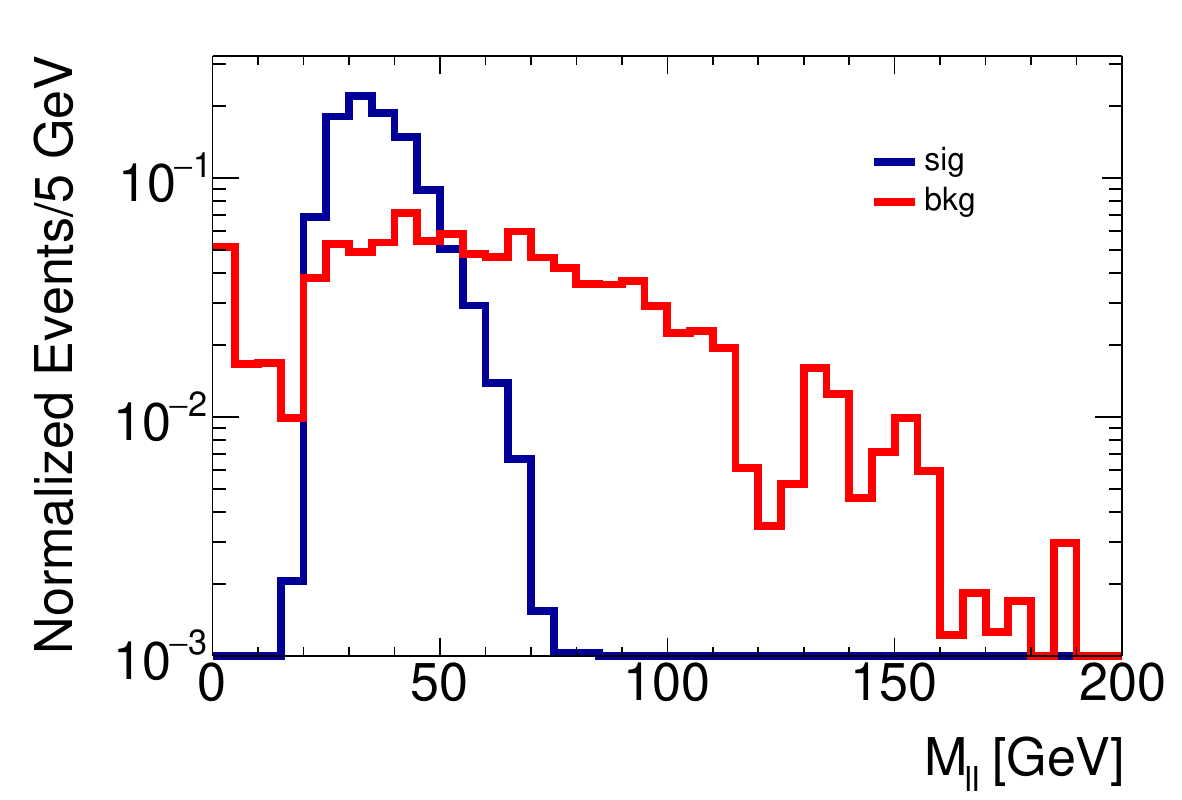}
	\end{minipage}
\caption{Distributions in invariant mass of lepton (left) and jet (right) pairs for BP2 and backgrounds  when $\cmsfourteen$  and L = $\threefbm$.}\label{f:MllMjj}
\end{figure}

In order to further separate signal from background events, we can also combine the two leptons and two jets together, denoting their invariant masses as $M_{ll}$ and $M_{jj}$, respectively, as shown in Fig.~\ref{f:MllMjj}. Since, for  signal events, all particles come from the SM-like Higgs boson decaying, any such combination only yield a smaller invariant mass, typically, about half of the SM-like Higgs boson mass. In contrast, for background events, these lepton and jet pairs always come from a heavy resonance, a top (anti)quark and a $W^\pm$ or $Z$ boson, so the background spectra are somewhat harder in general.

\begin{figure}[H]
	\centering
	\begin{minipage}{8cm}
		\centering
		\includegraphics[width=8cm]{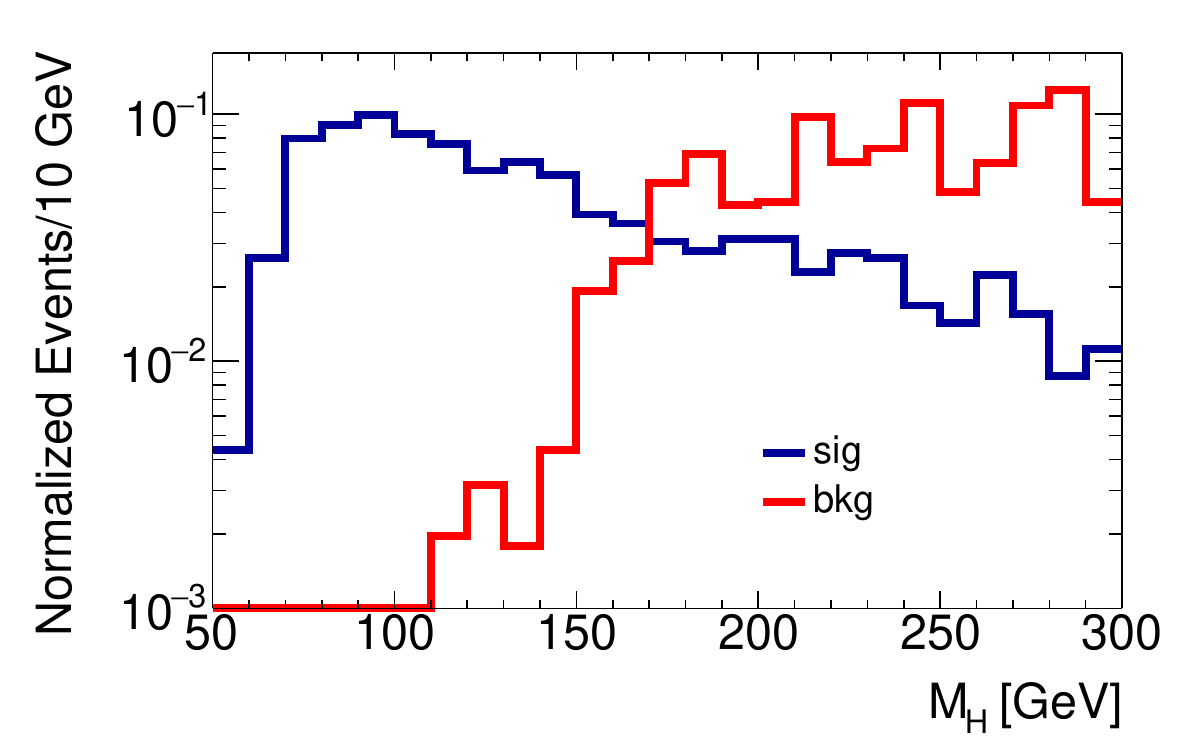}
	\end{minipage}
    \begin{minipage}{8cm}
		\centering
		\includegraphics[width=8cm]{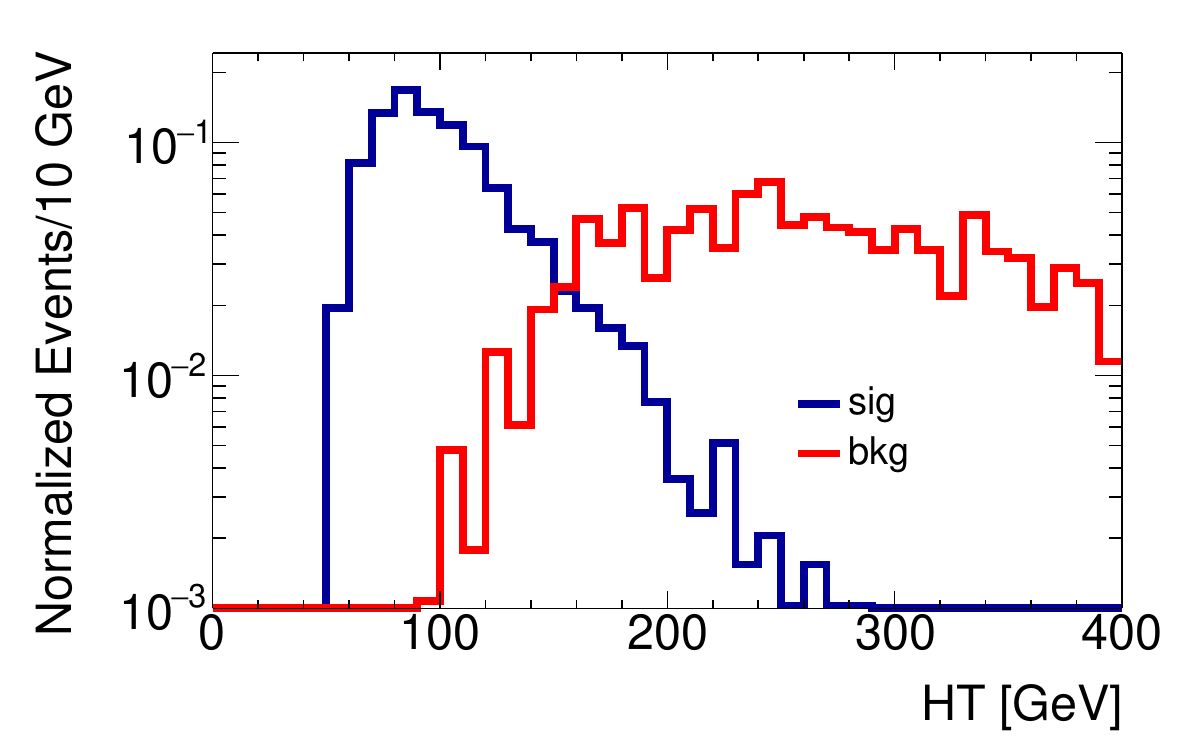}
	\end{minipage}
\caption{Distributions in invariant mass of two leptons and two jets (left) and in scalar sum of visible momenta (right) for BP2 and backgrounds  when $\cmsfourteen$  and L = $\threefbm$.}\label{f:MHHT}
\end{figure}

We can further reconstruct the invariant mass of the SM-like Higgs boson  $M_H$ with two leptons and two jets but without any MET. It is expected that the sampled values of it will be lower than the real SM-like Higgs boson mass. Nevertheless, it is helpful enough to distinguish the signal and background events further, see Fig.~\ref{f:MHHT} (left). 

Another useful kinematic variable is $H_T$, which is the scalar sum of all visible final states transverse momenta, which are shown in Fig.~\ref{f:MHHT} (right).

\subsection{Multi-variate analysis}

To optimize the signal and background distinction, a  Gradient-Boosted Decision Tree (GBDT) approach is further applied, which is implemented in the Toolkit for Multi-Variate Analysis (TMVA) within ROOT~\cite{Therhaag:2010zz}. 

\begin{figure}[ht]
\begin{minipage}{0.47\textwidth}
      \begin{center} 
                     (a)	\\
         \includegraphics[height=5.0cm]{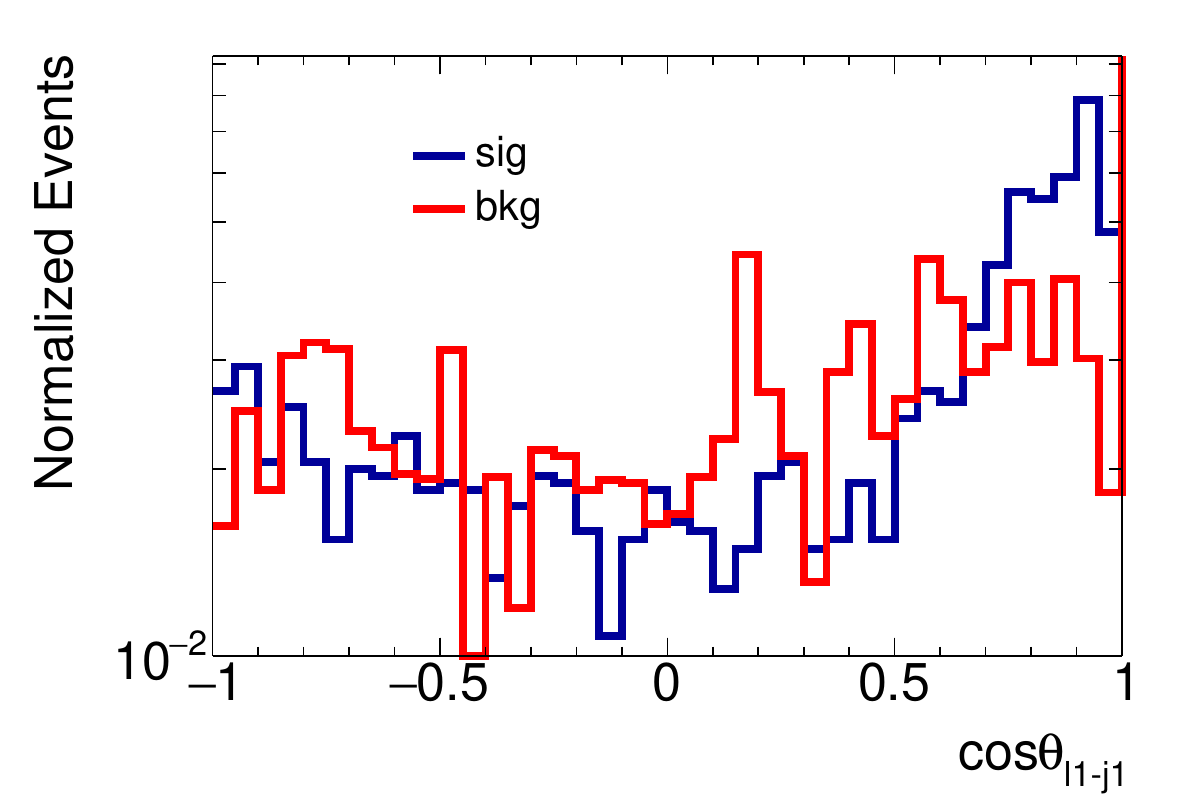}	
      \end{center}
\end{minipage}
\begin{minipage}{0.47\textwidth}
      \begin{center} 
                     (b)	\\
         \includegraphics[height=5.0cm]{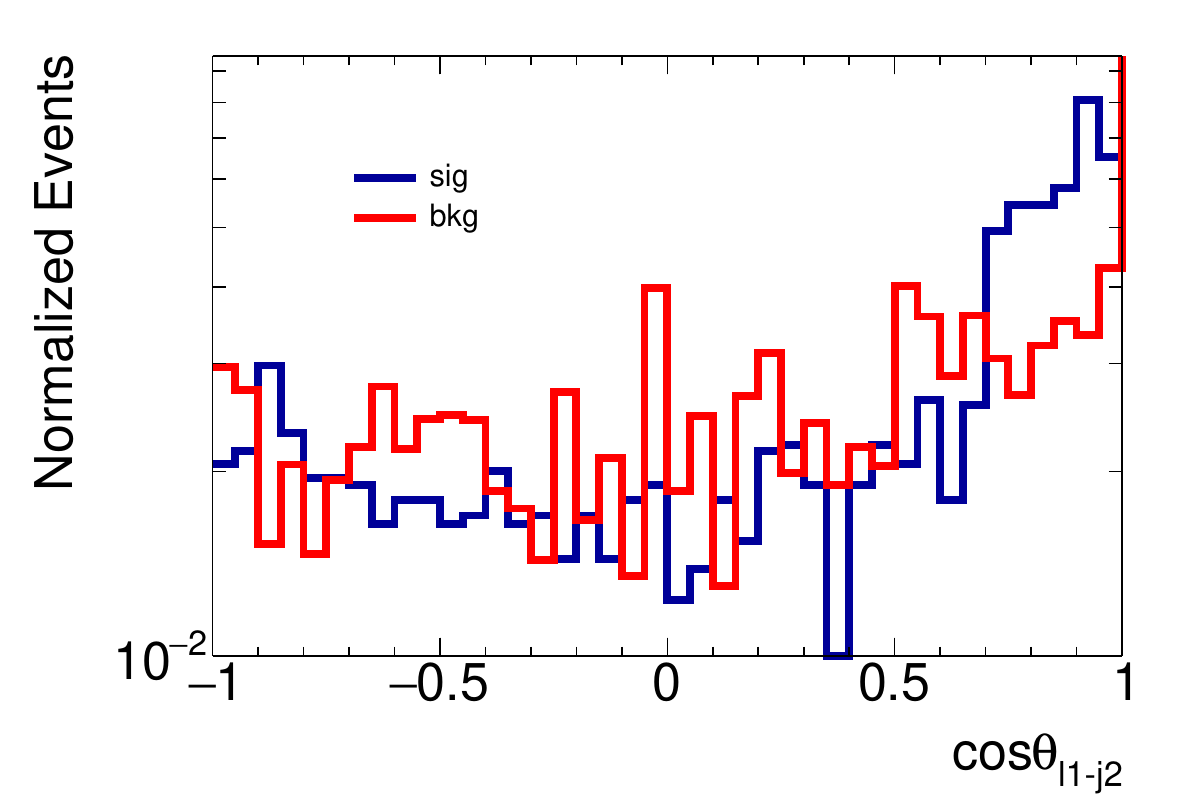}	
      \end{center}
\end{minipage}

\begin{minipage}{0.47\textwidth}
      \begin{center} 
                     (c)	\\
         \includegraphics[height=5.0cm]{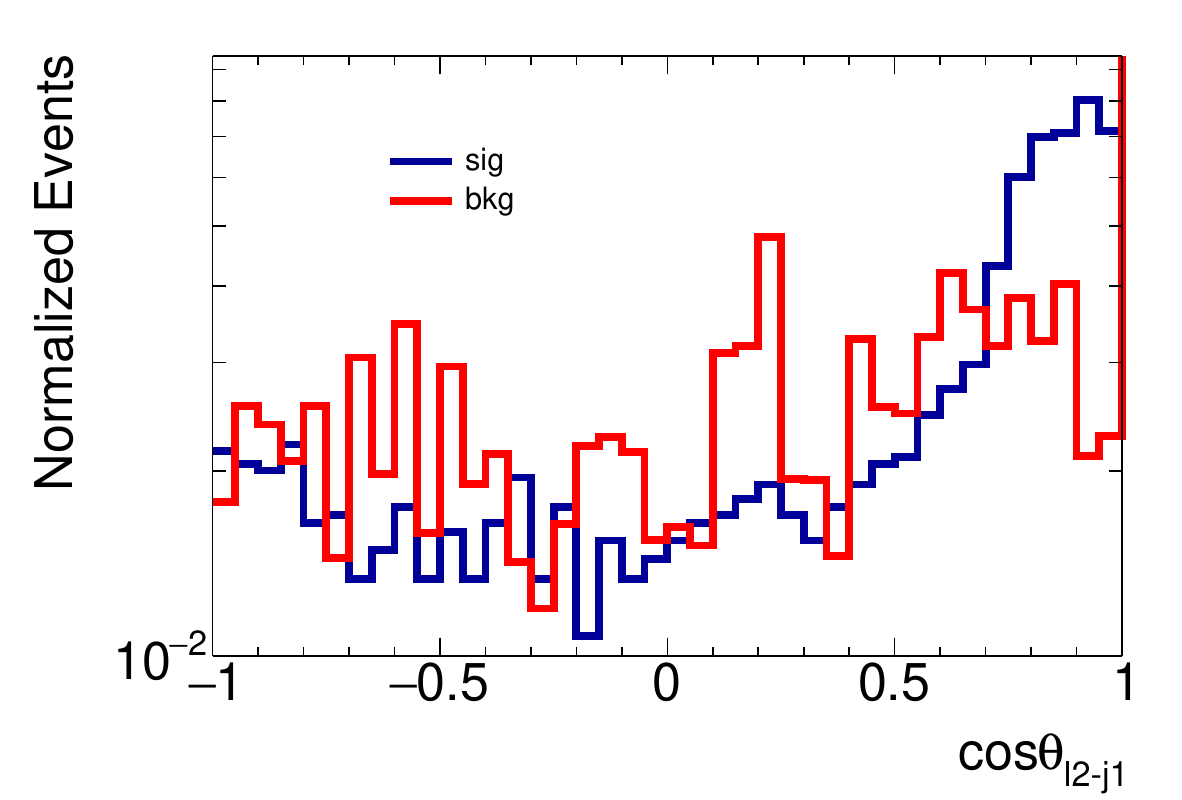}	
      \end{center}
\end{minipage}
\begin{minipage}{0.47\textwidth}
      \begin{center} 
                     (d)	\\
         \includegraphics[height=5.0cm]{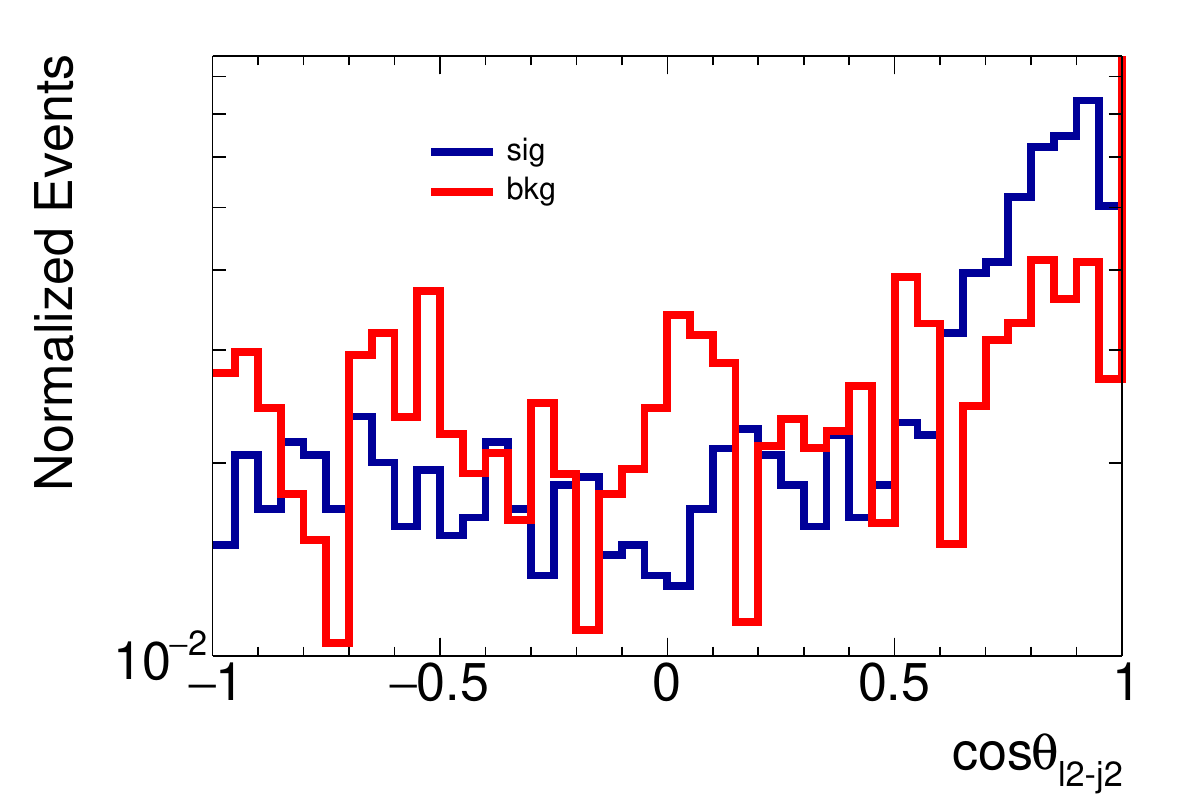}	
      \end{center}
\end{minipage}

\begin{minipage}{0.47\textwidth}
      \begin{center} 
                     (e)	\\
         \includegraphics[height=5.0cm]{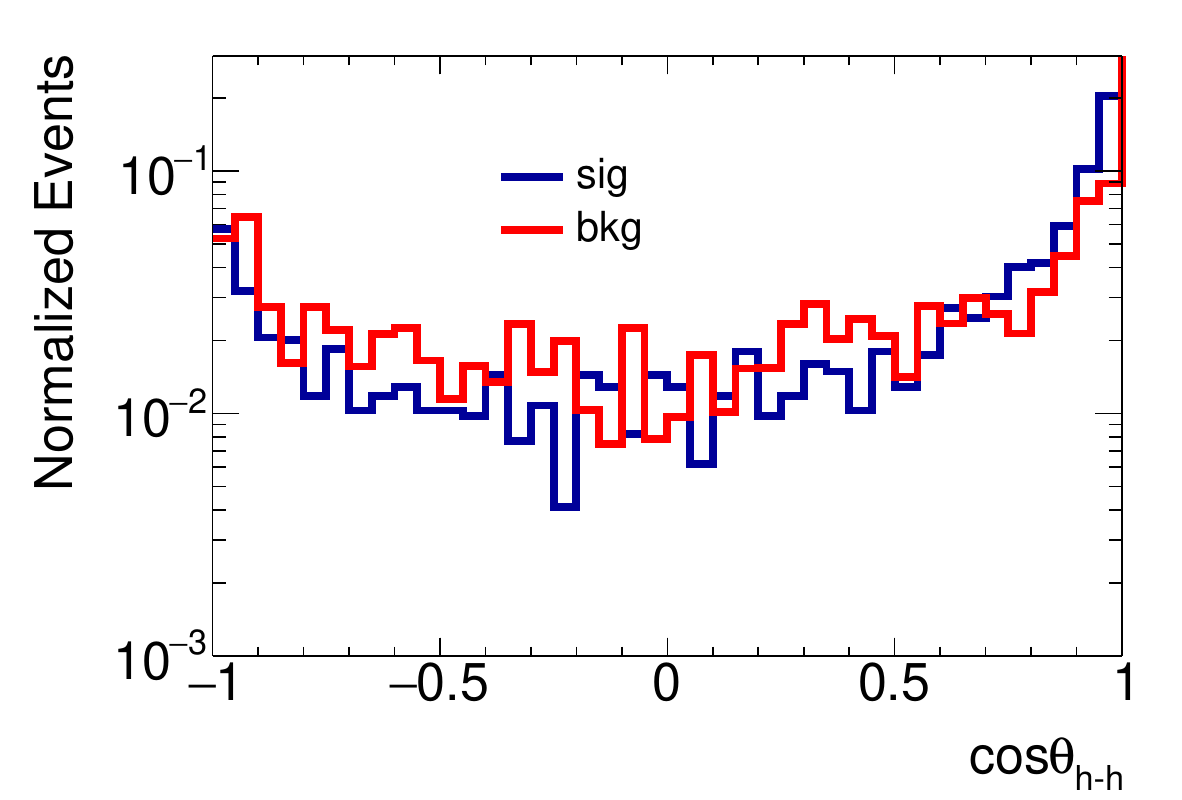}	
      \end{center}
\end{minipage}

 \caption{Distributions in the (cosine of the) angle between (a)-(d) pairs of jets/leptons and (e) the two reconstructed light Higgs states for BP2 and backgrounds  when $\cmsfourteen$  and L = $\threefbm$. 
 }\label{f:angle}
\end{figure}

Ten input variables in total are used for the GBDT/TMVA analysis, which are shown in Tab.~\ref{t:MVA_observables}. In addition to the above invariant masses and $H_T$, we also calculate five angles between pairs of objects in the final state, $\cos(\theta_{l_1j_1})$,  $\cos(\theta_{l_1j_2})$, $\cos(\theta_{l_2j_1})$, $\cos(\theta_{l_2j_2})$ and  $\cos(\theta_{h_1h_2})$, 
which is defined as the cosine of the angle between the two reconstructed final states states. The distributions are shown in Fig.~\ref{f:angle}.
As we expect these to be useful, since in the signal they all tend to be collinear to each other (because of the boost structure herein) unlike the background case. 

\begin{table}[H]
 \begin{center}
 \begin{small}
\centerline{}
\begin{tabular}{|c|c|c|c|c|c|}
\hline
Energy variables& $M_{l\tau_{j}}^{1}$& $M_{l\tau_{j}}^{2}$ & $M_{ll}$ & $M_{jj}$ & $H_T$\\
\hline
Angular variables &$\cos(\theta_{h_1h_2})$ & $\cos(\theta_{l_1j_1})$ & $\cos(\theta_{l_1j_2})$ &$\cos(\theta_{l_2j_1})$ & $\cos(\theta_{l_2j_2})$\\
\hline
\end{tabular}
 \end{small}
   \caption{The input observables used in the GBDT analysis.}\label{t:MVA_observables} 
 \end{center}
  \end{table} 

\begin{figure}[H]
	\centering
		\centering
		\includegraphics[width=7cm]{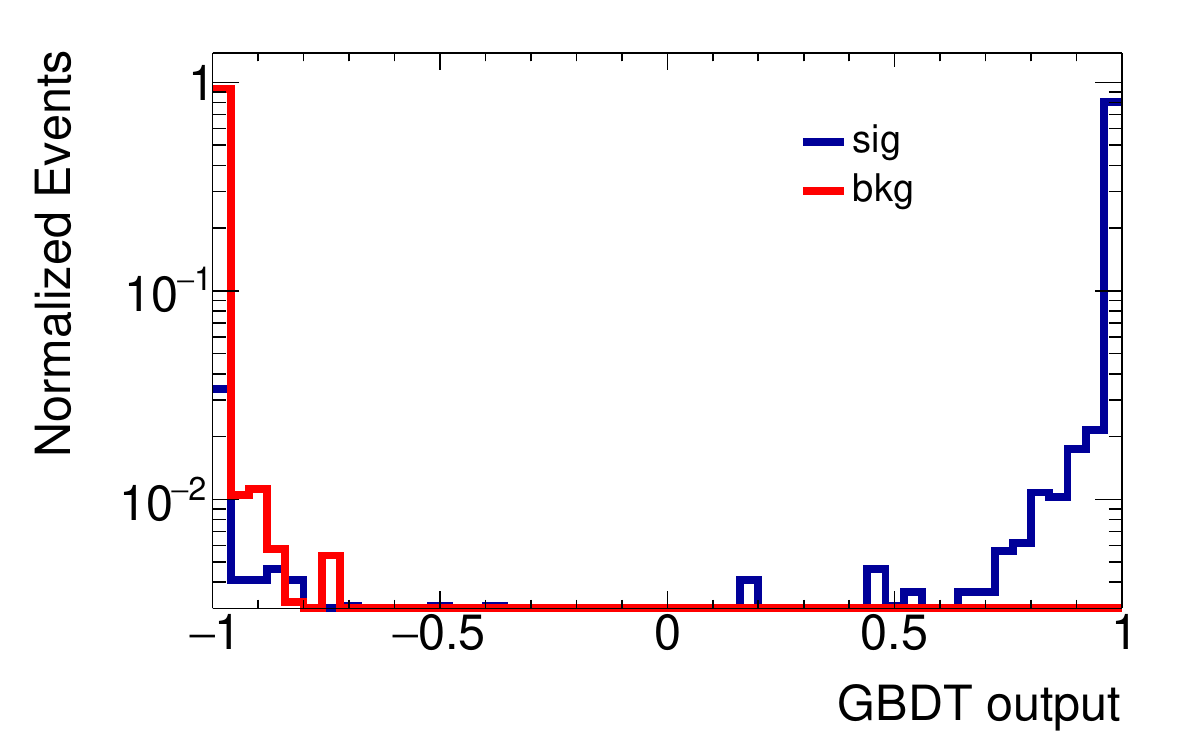}
\caption{GBDT output for the BP2 signal and backgrounds. 
}\label{f:MVA}
\end{figure}

In the training stage, it is found that the invariant mass variables are powerful, but angular variables are still useful. Ultimately, for the GBDT output, a very good separation between the signal and backgrounds can be achieved, which is shown in Fig.~\ref{f:MVA}. 

\subsection{Significances}

The process of searching for signals at high energy colliders using ML techniques is ultimately a statistical process. Therefore, to look for signals, a large number of  events need to be collected and the kinematic variables of the final state particles need to be analyzed to find differences from the noise, but then a sizeable statistical sample needs to be collected, so has to have enough significance. 


The latter
is defined as $\Sigma=\frac{N_{S}}{\sqrt{(N_{S} + N_{B})}}$, where $N_{S}$ and $N_{B}$ represent the number of events for the signal ($B$) and (total) background ($B$), respectively, at a given luminosity $L$.

To distinguish the signal from the background and to compute the ensuing significance, we apply six kinematic cuts on reconstructed observables and one MVA cut on GBDT outputs, the details of which are listed in Tab.~\ref{t:BP2}, for BP2 and where $L=\threefbm$ (with a CM energy $\cmsfourteen$, as previously mentioned). Following the acceptance cuts and upon applying the SS lepton criterion (the combination of which we refer to as pre-selection), we see a strong decrease of the initially huge SM backgrounds to the same order as the signal. For example, the initially troublesome $Zjj$ is almost eliminated after this cut (thus, it is no longer included in this table). As Figs.~\ref{f:Mh1Mh2}--\ref{f:MHHT} show, the kinematic selection is already quite efficient is
separating signal from background, as -- depending on the BP adopted -- $\Sigma$ values can reach values around 4, even with rather loose cuts. However, exploitation of the GBDT output more than  double the final significance.

\begin{table}[H]
	\begin{center}
 \begin{tabular}{|c| c| c| c| c| c| c| c| c| c| c| c| c|}
 \hline
 $                  L  =\threefbm  $ &  $S$ &  $      t\bar{t} $ &  $   t\bar{t}W^+W^- $ &  $     t\bar{t}Z $ &  $   t\bar{t}ZZ $ &  $        Wtb $ &  $     W^+W^-jj $ &  $        ZZ $ &                $B$ &                          $\Sigma$\\ 
 \hline
Acceptance &                    262.6 &              2.17$\times10^{6}$ &                      2.3 &                       63 &                     0.01 &                  53751.6 &                   119699 &                   2790.8 &              2.35$\times10^{6}$ &                                  0.17\\ 
 \hline
 \hline
SS Leptons &                    131.5 &                   5829.8 &                      1.5 &                     23.6 &                     0.01 &                    186.6 &                    595.5 &                    754.1 &                   7391.1 &                     1.52\\ 
 \hline
 $                      H_T \in [ 40 ,200 ] $ &                    129.9 &                   1040.7 &                    0.002 &                      0.5 &                    1e-05 &                     40.4 &                    127.9 &                    379.1 &                   1588.7 &                     3.13\\ 
 \hline
 $                   M_{ll} \in [ 10 ,80 ] $ &                    129.8 &                    905.8 &                    0.001 &                      0.5 &                    1e-05 &                     32.7 &                     58.4 &                    283.8 &                   1281.1 &                     3.45\\ 
 \hline
 $                M_{h_{1}} \in [ 0 ,150 ] $ &                    128.7 &                    809.4 &                    0.001 &                      0.4 &                    1e-05 &                     31.1 &                     56.9 &                    271.4 &                   1169.2 &                     3.57\\ 
 \hline
 $                M_{h_{2}} \in [ 0 ,150 ] $ &                    126.9 &                    732.3 &                    0.001 &                      0.4 &                    8e-06 &                       28 &                     55.3 &                    260.5 &                   1076.5 &                     3.66\\ 
 \hline
 $                   M_{H} \in [ 20 ,230 ] $ &                     96.9 &                    269.8 &                        0 &                      0.2 &                    5e-06 &                     10.9 &                     26.9 &                    124.3 &                      432 &                     4.21\\ 
 \hline
 $                      {\rm GBDT} \in [ 0.5 ,1 ] $ &                     92.6 &                        0 &                        0 &                     0.02 &                        0 &                        0 &                      3.2 &                      7.1 &                     10.2 &                     9.13\\ 
 \hline
 \end{tabular}
 \caption{Response to our selection cuts for the signal  (e.g., BP2) and background (separately and  total) rates computed at $\cmsfourteen$  for $\threefbm$. }\label{t:BP2} 
 \end{center}
 \end{table}

The results for all twelve BPs are shown in Tab.~\ref{t:significance}. With only the acceptance and SS lepton criteria, the significances can reach 3 for BP8, BP9 and BP11. After applying the kinematic cuts, the significances can be all larger than 3, except for BP3. Upon applying the MVA selection, significances can reach values as large as 14 for BP8. In short, it is clear that a judicious combination of pre-selection and kinematic cuts combinaed with ML analysis can result in large sensitivity to a significant expanse of 2HDM Type-X parameter space through our chosen signal.

\begin{table}[H]
	\begin{center}
			\begin{tabular}{|c| c| c| c| c| c| c| c| c| c| c| c| c|}
				\hline
			     $\Sigma$        & BP1  & BP2   & BP3   & BP4  & BP5   & BP6   & BP7   & BP8   & BP9   & BP10  & BP11   & BP12 \\
                    \hline
    after selecting SS leptons    & 1.17 &  1.52 & 1.06  & 1.40 &  2.04 &  1.79 &  1.71 &  3.89 &  3.15 &  2.86 &  3.18  &  2.58\\
                    \hline
    after kinematic cuts, w/o GBDT & 3.09 &  4.21 & 2.66  & 3.79 &  5.20 &  4.50 &  4.50 &  8.84 &  7.58 &  6.65 &  7.41  &  6.24\\
				\hline
    after kinematic cuts, w/ GBDT  & 7.65 &  9.13 & 6.27  & 8.61 & 10.41 &  9.45 &  9.34 & 14.32 & 12.91 & 11.25 & 12.14  & 10.74\\
				\hline 
			\end{tabular}
			\caption{Significances following our different selections for all signals (BP1--BP12) at $\cmsfourteen$ for $\threefbm$.
            }\label{t:significance}
	\end{center}
\end{table}


\section{Conclusions}
In this paper, we have investigated the feasibility of producing the SM-like Higgs boson ($H$) in single mode  at the LHC  decaying via a resonance into two lighter neutral CP-even Higgs bosons ($hh$) in the 2HDM Type-X.  We have focused on $4\tau$ final states, where two $\tau$'s decay leptonically to form a pair of SS leptons (electrons and/or muons) and two $\tau$'s decay hadronically, thus producing two jets. In order to motivate future studies of this signature, we have also proposed  12 BPs and processed these through a realistic MC analysis for a CM (proton-proton) energy $\cmsfourteen$ and (integrated) luminosity $L = \threefbm$. Eventually, we have proven that this kind of signal could be found at the end of Run 3 of the LHC, following  a dedicated selection based on kinematic and TMVA analysis.

\section*{Acknowledgments}
 SM is supported in part through the NExT Institute and the STFC Consolidated
Grant ST/L000296/1. SS is supported in full through the NExT Institute and
acknowledges the use of the IRIDIS High Performance Computing Facility, and associated
support services, at the University of Southampton, in the completion of this work. YW’s
work is supported by the Natural Science Foundation of China Grant No. 12275143, the Inner
Mongolia Science Foundation Grant No. 2020BS01013 and the Fundamental Research Funds
for the Inner Mongolia Normal University Grant No. 2022JBQN080. QSY is supported by the
Natural Science Foundation of China under the Grants No. 11875260 and No. 12275143.

\end{document}